\documentclass{iopconfser}
\pdfoutput=1

\usepackage{bm}
\usepackage[colorlinks=true, urlcolor=blue, linkcolor=blue, citecolor=blue, pdftex]{hyperref}
\usepackage{float}
\usepackage[usenames,dvipsnames]{xcolor}
\usepackage{comment}
\usepackage[utf8]{inputenc}
\usepackage{braket}
\usepackage{graphicx}
\usepackage[numbers,sort&compress]{natbib}
\usepackage{subfigure}
\usepackage{dcolumn}
\usepackage{amsmath,amssymb}
\usepackage{amsthm}
\usepackage{mathrsfs}
\usepackage{indentfirst}
\usepackage{enumerate}
\usepackage{fancyhdr}
\usepackage{caption}
\usepackage{tikz}
\usetikzlibrary{arrows, shapes.gates.logic.US, calc}
\usetikzlibrary{angles,quotes}
\tikzstyle{branch}=[fill, shape=circle, minimum size=3pt, inner sep=0pt]
\usepackage{mathtools}
\usepackage{color,graphicx} 
\usepackage{bbold}
\usepackage{dsfont}

\usepackage{hyperref}
\hypersetup{
 colorlinks=true,
 citecolor=blue,
 linkcolor=blue,
 urlcolor=blue,
 pdfpagemode=UseNone,
 pdfstartview=FitH}

\usepackage{subfiles}

\definecolor{C0}{HTML}{1f77b4}
\definecolor{C1}{HTML}{ff7f0e}
\definecolor{C2}{HTML}{2ca02c}
\definecolor{C3}{HTML}{d62728}

\begin{document}
\title{Equivariant Variational Quantum Eigensolver to detect Phase Transitions through Energy Level Crossings}

\author{Giulio Crognaletti$^{1,2,*}$, Giovanni Di Bartolomeo$^{1,2}$, Michele Vischi$^{1,2}$, Luciano Loris Viteritti$^{1}$}

\affil{$^1$University of Trieste, Trieste, Italy}
\affil{$^2$Istituto Nazionale di Fisica Nucleare, Trieste Section, Italy}

\affil{$^*$Corresponding author}

\email{giulio.crognaletti@phd.units.it, giovanni.dibartolomeo@phd.units.it, michele.vischi@phd.units.it,  lucianoloris.viteritti@phd.units.it}

\begin{abstract}
Level spectroscopy stands as a powerful method for identifying the transition point that delineates distinct quantum phases. Since each quantum phase exhibits a characteristic sequence of excited states, the crossing of energy levels between low-lying excited states offers a reliable mean to estimate the phase transition point. While approaches like the Variational Quantum Eigensolver are useful for approximating ground states of interacting systems using quantum computing, capturing low-energy excitations remains challenging. In our study, we introduce an equivariant quantum circuit that preserves the total spin and the translational symmetry to accurately describe singlet and triplet excited states in the $J_1$-$J_2$ Heisenberg model on a chain, which are crucial for characterizing its transition point. Additionally, we assess the impact of noise on the variational state, showing that conventional mitigation techniques like Zero Noise Extrapolation reliably restore its physical properties.
\end{abstract}

\section{Introduction}\label{intro}
Exploring phase transitions is crucial for understanding the fundamental behavior of matter~\cite{Sachdev_1999}. These transitions are usually identified by analyzing order parameters that describe specific phase changes.
However, accurate estimates of transition points require large systems approaching the thermodynamic limit, which typically is highly computationally expensive.
An alternative approach involves utilizing level spectroscopy~\cite{eggert1996,sandvik2010,nomura1995}. 
In this framework, the ground state phase transition is detected by the energy level crossings of low-lying excited states. Indeed, if the quantum numbers of the ground state do not change across the phase transition, but those of the low-energy excited states do change, then a correspondence between excited energy level crossings and the phase transitions is expected~\cite{nakamura1999,nakamura2000,sandvik2004,sandvik2010b}. For several one-dimensional models this mapping is well established. Moreover, it seems to be a promising approach also for two-dimensional spin systems~\cite{wang2018,ferrari2020,nomura2021}. The benefit of this method lies in its insensitivity to system size, enabling highly accurate predictions of the phase transition point even with small clusters.

In this work, we introduce a variational Ansatz adequate for the implementation on quantum devices to detect the transition by exploiting level spectroscopy. In particular, we focus on the $J_1$-$J_2$ Heisenberg model on a chain, where level crossing has been successfully applied with classical methods. The Hamiltonian of the model reads
\begin{equation}\label{eq:J1J2_ham}
    \hat{H}_{J_1\text{-}J_2} = J_1\sum_{r=1}^N \hat{\boldsymbol{S}}_{r}\cdot\hat{\boldsymbol{S}}_{r+1} + J_2\sum_{r=1}^N \hat{\boldsymbol{S}}_{r}\cdot\hat{\boldsymbol{S}}_{r+2} \ ,
\end{equation}
where $\hat{\boldsymbol{S}}_r = (\hat{S}_r^x,\hat{S}_r^y,\hat{S}_r^z)$ is the spin $1/2$ operator at site $r$ and $J_1, J_2 \ge 0$ are the antiferromagnetic couplings between nearest and next-nearest neighbors sites, respectively. For $J_2 = 0$ the model in Eq.~\eqref{eq:J1J2_ham} reduces to the one-dimensional Heisenberg model and the ground-state properties can be computed using the Bethe Ansatz~\cite{franchini2017}. However, when $J_2 > 0$ there are no exact solutions. Still, its phase diagram is well established by numerical and analytical calculations~\cite{white1996,lacroix2011}. 
In particular, a Berezinskii–Kosterlitz–Thouless (BKT) transition~\cite{jos2013} at $(J_2/J_1)_c = 0.24116(7)$ separates a gapless region at small values of $J_2/J_1$ by a gapped state at large values of the frustration ratio. On the one hand, the detection of a BKT transition is especially complicated by the direct computation of an order parameter. Indeed, large size effects are present and a huge number of sites is necessary to give a meaningful estimation of the phase transition point~\cite{white1996,viteritti2022,lacroix2011,viteritti2023}. On the other hand, by employing level spectroscopy, a small number of sites ($N \sim 30$) is sufficient to achieve a very accurate estimation~\cite{eggert1996}. 

As described earlier, this technique exploits excited states to determine the transition point. The eigenstates of the Hamiltonian in Eq.~\eqref{eq:J1J2_ham} can be classified according to the total spin $S$ and the momentum $k$, respectively the quantum numbers of $\boldsymbol{\hat{{S}}^2}$, where $\hat{\boldsymbol{S}}=\sum_{r=1}^N \hat{\boldsymbol{S}}_r$, and of the one-site translation operator $\hat{\mathcal{T}}$ (assuming periodic boundary conditions).
The transition point can be detected as the energy level crossing between the first singlet ($S=0$) and triplet ($S=1$) excited states with momentum $\pi$ with respect to the ground state. Defining variational states which are eigenstates of the total spin operator $\boldsymbol{\hat{S}^2}$ requires considerable computational effort when using classical methods like Tensor Networks and Neural-Network Quantum States (NQS). On the one hand, for NQS the wave function amplitude is typically represented in the basis along the $z$-direction, which inherently breaks total spin symmetry. To address this, it is necessary to define the amplitude in a spin-symmetric basis, as proposed in Refs.~\cite{vieijra2020,vieijra2021}. On the other hand, Tensor Networks can be constructed with $SU(2)$-symmetric tensors, allowing for a larger bond dimensions than what is possible when only $U(1)$ symmetry is imposed~\cite{singh2012,schmoll2020, yang2022, gong2014}. However, due to the complexity of incorporating non-abelian symmetries, both Tensor Network and NQS approaches typically restrict computations to the $S^z$ symmetry sectors, where highly accurate results can still be achieved~\cite{carleo2017,nomura2021,nomura2021b,roth2021,roth2023, viteritti2023, viteritti2024, rende2023, hibat2020}.

Conversely, preparing translational- and spin-equivariant quantum circuits can be achieved implementing only constant depth layers (see Sec.~\ref{sec:vqe}). Notably, combining the latter approach with the level spectroscopy technique on small clusters, makes the problem suitable for quantum computers, even in the near-term.  In this scenario, we devise an equivariant variational quantum circuit~\cite{Lyu_2023,gard2020efficient,Seki2020,Meyer2023,Chang2023,Le2023} preserving both total spin and translational symmetry, optimized through the Variational Quantum Eigensolver (VQE)~\cite{cerezo2021variational}. The latter is a hybrid quantum-classical algorithm based on the variational principle. It leverages quantum computers to create a parameterized variational state and measure its corresponding energy, while a classical loop iteratively updates its parameters to approach the minimum of the energy landscape. 
Importantly, the variational principle can be extended beyond approximating ground state properties to describe low-lying excited states as well. For instance, this can be achieved by constructing suitable variational states where specific symmetry sectors can be specified~\cite{Kattemolle_2022,mizusaki2004,viteritti2022,nomura2021,nomura2021b}. In this work, we explore this possibility through numerical simulations of ideal quantum circuits, focusing on the variational approximation of the excited states of the $J_1$-$J_2$ Heisenberg model in Eq.~\eqref{eq:J1J2_ham}. However, our approach can be suitably extended to handle other Hamiltonians, and it results particularly valuable to treat those that conserve total spin.

The conceptual aim of this work is to demonstrate that condensed matter techniques, such as level spectroscopy, can be combined with quantum algorithms to detect quantum phase transitions. While previous works~\cite{Lyu_2023, gard2020efficient} discussed the construction of quantum circuits that respect certain symmetries, we take an additional step by showing that the knowledge of specific excited states can be leveraged to detect phase transitions that are challenging to characterize by directly measuring order parameters due to significant finite-size effects. From a methodological standpoint, we also introduce a general approach based on the \textit{Linear Combination of Unitaries} (LCU) for a posteriori symmetrization. This technique can complement symmetric Ans\"atze by restoring symmetries that are difficult to encode directly into the circuit design.

Moreover, as quantum hardware is highly susceptible to the interaction with the environment, understanding the effects of noise in quantum circuits becomes crucial~\cite{preskill2018quantum}.
For this reason, the sensitivity of our approach to noise is evaluated through numerical simulations of noisy quantum circuits. Despite the explicit breaking of variational state symmetries in the presence of noise~\cite{Tüysüz2024}, we demonstrate that by integrating standard error mitigation techniques, such as Zero Noise Extrapolation (ZNE)~\cite{PhysRevLett.119.180509}, we can successfully restore the desired physical properties.

\section{Variational Ansatz}\label{vqe}
\subsection{Variational Quantum Circuit}\label{sec:vqe}

Given a Hamiltonian $\hat{H}$, the VQE offers a method to approximate its eigenstates by exploiting the variational principle, which involves minimizing the variational energy $E_\theta = {\bra{\Psi_\theta} \hat{H} \ket{\Psi_\theta}}$. Here, $\ket{\Psi_\theta}$ constitutes a variational state depending on $\theta$, a vector of parameters. 
In the VQE framework, the Ansatz is represented as a quantum circuit identified by a unitary transformation $\hat{U}_{\theta}$ acting on an initial state $\ket{\phi}$, such that $\ket{\Psi_{\theta}} = \hat{U}_{\theta}\ket{\phi}$. In this study, we utilize a quantum circuit based on the Hamiltonian Variational Ansatz (HVA)~\cite{cerezo2021variational,wecker2015progress}, which proved to be effective for approximating quantum many-body eigenstates~\cite{wiersema2020,mele2022,anselme2022,wierichs2020,ho2019,Seki2020,feulner2022}.
The HVA approach involves introducing a set of auxiliary Hamiltonians $\hat{H}_1, \hat{H}_2, \dots, \hat{H}_{M}$ such that $\left[ \hat{H}_{m}, \hat{H}_{m'} \right] \neq 0 \; \forall m \neq m'$. The variational state is then expressed as:
\begin{equation}
    \ket{\psi_{\theta}} = \prod_{l = 1}^{L} e^{-i\theta_{M}^{l}\hat{H}_M}
    \dots e^{-i\theta^{l}_{2}\hat{H}_2}  e^{-i\theta^{l}_{1}\hat{H}_1}\ket{\phi} \ ,
\end{equation}
where the initial state $\ket{\phi}$ is typically identified as a low-energy eigenstate of one of the auxiliary Hamiltonians $\hat{H}_m$ ($m>1$).
The number of variational parameters in the HVA approach is $M \cdot L$ and the accuracy of the variational state is mainly determined by the number of layers $L$.

\begin{figure}[t]
    \center
     \includegraphics[width=\columnwidth]{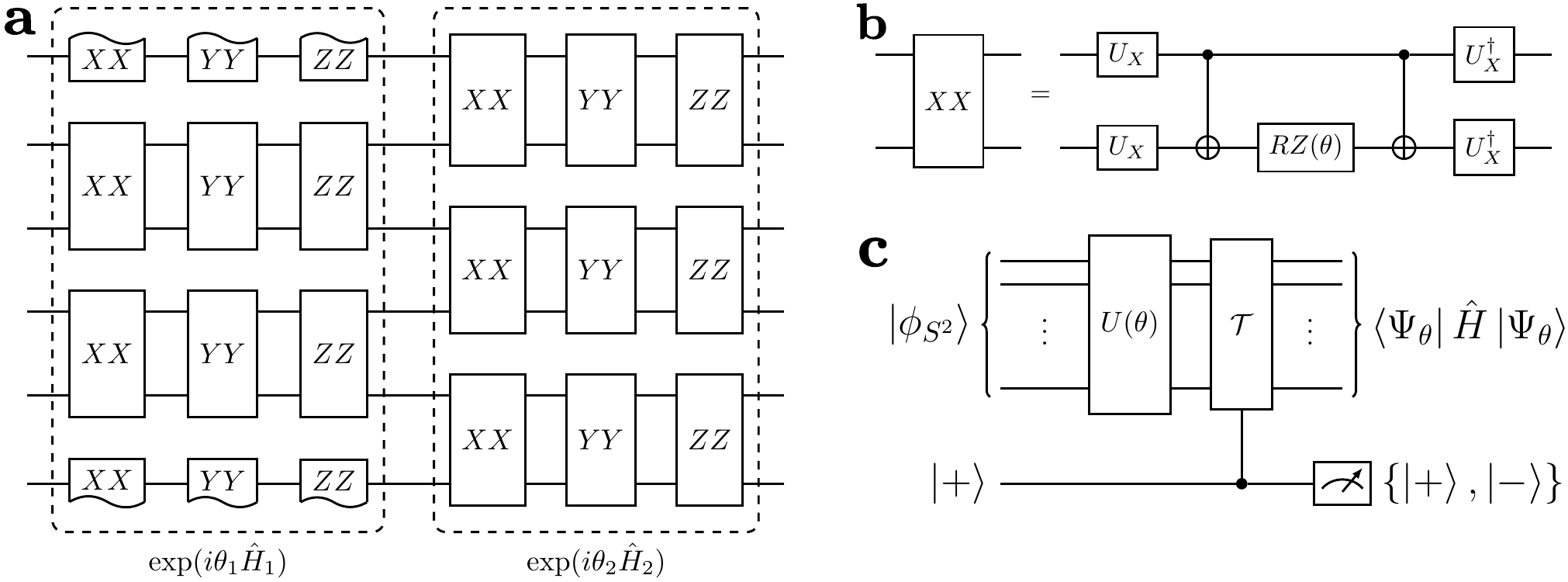}
    \caption{\textbf{a)} Implementation of a single layer of the quantum circuit defined in Eq.~\eqref{eq:unitary3} of the main text. This circuit consists of 2-local rotations, such as $XX$, which denotes the unitary $\exp(i\theta \hat{S}^x_r \otimes \hat{S}^x_{r+1})$ and similarly for $YY$ and $ZZ$. \textbf{b)} Implementation of the 2-local rotation $XX$. It includes two CNOTs, a $RZ$ rotation and diagonalizing gates $U_X$, defined by $\hat{U}^\dagger_X \hat{S}^x_r \hat{U}_X = \hat{S}^z_r$. The implementation is analogous for the gates $YY$ and $ZZ$. \textbf{c)} Circuit used to perform symmetrization by LCU, where a controlled $\hat{\mathcal{T}}$ is added to $\hat{U}_\theta$ to symmetrize $\ket{\Psi_\theta}$.}
    \label{fig:circuits}
\end{figure}

In the HVA framework, the standard practice is to require that $\sum_{m=1}^M \hat{H}_m = \hat{H}$. However, we relax this constraint while still maintaining a connection between $\sum_{m=1}^M \hat{H}_m$ and $\hat{H}$. Indeed, our focus lies on the symmetry properties of the auxiliary Hamiltonians.

The simplest way to define a non-trivial variational state $\ket{\Psi_\theta}$ is to consider a set of $M=2$ auxiliary Hamiltonians $\hat{H}_1$ and $\hat{H}_2$. As a result, the unitary operator takes the following form:
\begin{equation}\label{eq:unitary}
    \hat{U}_{\theta} = \prod_{l=1}^{L} e^{i \theta^l_2 \hat{H}_{2}}e^{i \theta^l_1 \hat{H}_{1}} \ .
\end{equation}
For the $J_1$-$J_2$ Heisenberg model in Eq.~\eqref{eq:J1J2_ham}, one way to define $\hat{H}_1$ and $\hat{H}_2$ is to observe that the nearest-neighbor term ${\sum_{r} \hat{\boldsymbol{S}}_{r}\cdot\hat{\boldsymbol{S}}_{r+1}}$ can be decomposed into the sum of two contributions
${\hat{H}_{\text{even}}=\sum_{r=1}^{N/2} \hat{\boldsymbol{S}}_{2r-1}\cdot\hat{\boldsymbol{S}}_{2r}}$ and ${\hat{H}_{\text{odd}}=\sum_{r=1}^{N/2} \hat{\boldsymbol{S}}_{2r}\cdot\hat{\boldsymbol{S}}_{2r+1}}$. Additionally, the ground state of $\hat{H}_{\text{even}}$ is a product state of singlet pairs, and its excited states can be constructed by replacing a singlet pair with a triplet one (see Sec.~\ref{sec:symmetries}). Therefore, we set $\hat{H}_2 = \hat{H}_{\text{even}}$, and choose the state $\ket{\phi}$ to be an appropriate eigenstate of $\hat{H}_{\text{even}}$. Subsequently, the most natural choice for the other term is $\hat{H}_1 = \hat{H}_{\text{odd}}$, since $[\hat{H}_{\text{even}},\hat{H}_{\text{odd}}]\neq 0$ (see Sec.~\ref{sec:init_ham} of the \textit{Appendix} for a discussion about an alternative choice of $\hat{H}_1$ and $\hat{H}_2$).
Since the Hamiltonians $\hat{H}_{1}$ and $\hat{H}_{2}$ are defined as sums of commuting terms, the unitary operator $\hat{U}_{\theta}$ in Eq.~\eqref{eq:unitary} becomes
\begin{equation}\label{eq:unitary2}
    \hat{U}_{\theta} = \prod_{l=1}^{L} \prod_{r=1}^{N/2}e^{i \theta^l_2 \hat{\boldsymbol{S}}_{2r-1}\cdot\hat{\boldsymbol{S}}_{2r}} \prod_{q=1}^{N/2} e^{i \theta^l_1 \hat{\boldsymbol{S}}_{2q}\cdot\hat{\boldsymbol{S}}_{2q +1}} \ .
\end{equation}
Furthermore, considering the definition ${\hat{\boldsymbol{S}}_{2r-1} \cdot \hat{\boldsymbol{S}}_{2r} = \sum_{\alpha}\hat{S}_{2r-1}^{\alpha}\hat{S}_{2r}^{\alpha}}$ where $\alpha = x, y, z$, it is worth noting that $[\hat{S}_{2r-1}^{\alpha}\hat{S}_{2r}^{\alpha}, \hat{S}_{2r-1}^{\beta}\hat{S}_{2r}^{\beta}]=0$, and similarly for $\hat{\boldsymbol{S}}_{2r} \cdot \hat{\boldsymbol{S}}_{2r+1}$. Consequently, we can rewrite Eq.~\eqref{eq:unitary2} without any approximation:
\begin{equation}\label{eq:unitary3}
    \hat{U}_{\theta} = \prod_{l=1}^{L} \prod_{r=1}^{N/2} \prod_{\alpha} e^{i \theta^l_2 \hat{S}^{\alpha}_{2r-1}\hat{S}^{\alpha}_{2r}} \prod_{q=1}^{N/2} \prod_{\beta}e^{i \theta^l_1 \hat{S}^{\beta}_{2q}\hat{S}^{\beta}_{2q +1}} \ .
\end{equation}

\noindent The last step is crucial as it enables the implementation of each layer of $\hat{U}_\theta$ in constant depth relative to the number of qubits $N$. In fact, $\hat{U}_\theta$ in Eq.~\eqref{eq:unitary3} is the composition of 2-local gates, many of which can be executed simultaneously (see Fig.~\ref{fig:circuits}a and Fig.~\ref{fig:circuits}b). Note that in Fig.~\ref{fig:circuits}b we show a straightforward implementation of the two local gates. Alternatively, optimized decompositions that reduce the number of required CNOT gates, as proposed in Refs.~\cite{Vatan2004,Lyu_2023}, can also be implemented.

\subsection{Symmetries of the variational state}\label{sec:symmetries}
An effective strategy for approximating excited states involves constructing variational states with definite quantum numbers, thereby energy minimizations are performed within specific symmetry sectors. In order to detect the phase transition point in the $J_1$-$J_2$ Heisenberg model, we need to fix both the momentum $k$ and the total spin $S$. As a result $\ket{\Psi_{\theta}}$ should be invariant with respect to the operators $\boldsymbol{\hat{S}^2}$ and $\hat{\mathcal{T}}$. Given $\ket{\Psi_{\theta}}=\hat{U}_{\theta}\ket{\phi}$, if the initial state $\ket{\phi}$ lies in a definite symmetry sector and the unitary operator $\hat{U}_{\theta}$ preserves that symmetry (i.e., it is equivariant), then the resulting variational state $\ket{\Psi_{\theta}}$ can effectively estimate low-energy excitations.
Regarding the total spin, each rotation $e^{i\theta \hat{\boldsymbol{S}}_{j}\cdot\hat{\boldsymbol{S}}_{k}}$ in Eq.~\eqref{eq:unitary2} is equivariant under spin symmetry, namely $[\boldsymbol{\hat{S}^2}, e^{i\theta \hat{\boldsymbol{S}}{j}\cdot\hat{\boldsymbol{S}}{k}}] = 0$ $\forall j,k$. Therefore, to restrict the optimization to a specific spin sector, it suffices to choose $\ket{\phi}$ with a definite quantum number $S$.

The singlet state ($S = 0$) is implemented as the ground state of $\hat{H}_2$ 
\begin{equation}
\label{eq:singlet_state}
\ket{\phi_0} = \prod_{r=1}^{N/2} \ket{s}_{2r-1,2r} \ ,
\end{equation}
where $\ket{s}_{r,r'} = (\ket{0}_{r}\ket{1}_{r'}-\ket{1}_{r}\ket{0}_{r'})/\sqrt{2}$ is a singlet pair. Similarly, the triplet state ($S=1$) can be constructed as the first excited state of $\hat{H}_2$ by replacing in Eq.~\eqref{eq:singlet_state} a singlet pair with a triplet one $\ket{t}_{r,r'}=(\ket{0}_{r}\ket{1}_{r'}+\ket{1}_{r}\ket{0}_{r'})/\sqrt{2}$. Thus we can select as initial state either of the following $N/2$ degenerate states
\begin{equation}
\label{eq:triplet_state_tilde}
    \ket{\tilde{\phi}^j_1} = \ket{t}_{2j-1,2j}\prod_{r=1, r\neq j}^{N/2} \ket{s}_{2r-1,2r} \ ,
\end{equation}
where $j = 1, \dots ,N/2$.
Concerning translational invariance, while the Hamiltonian $\hat{H}_{J_1\text{-}J_2}$ in Eq.~\eqref{eq:J1J2_ham} preserves one-site translations, the unitary operator $\hat{U}_{\theta}$ does not. However, the Hamiltonians $\hat{H}_1$ and $\hat{H}_2$ are equivariant under translations of two lattice sites, meaning $[\hat{H}_1,\mathcal{\hat{T}}^2]=[\hat{H}_2, \mathcal{\hat{T}}^2]=0$. Consequently, by sharing the variational parameters across different qubits [see Eq.~\eqref{eq:unitary3}], we easily achieve $[\hat{U}_{\theta},\mathcal{\hat{T}}^2]=0$.
The latter condition leads us to consider a simple way to restore the one-site symmetry, which is necessary to fix the momentum $k$ in the variational state. This involves defining initial states $\ket{\phi}$ that possess translational invariance over two lattice sites. Notably, the state $\ket{\phi_0}$ in Eq.~\eqref{eq:singlet_state} already exhibits this symmetry, while the states $\ket{\tilde{\phi}_1^j}$ in Eq.~\eqref{eq:triplet_state_tilde} lack it. However, we can prepare a superposition of $\ket{\tilde{\phi}^j_1}$ states as 
\begin{equation}\label{eq:triplet_state}
    \ket{\phi_1} = \frac{1}{\sqrt{N/2}} \sum_{j=1}^{N/2} \ket{\tilde{\phi}^j_1} \ ,
\end{equation}
implying $\mathcal{\hat{T}}^2\ket{\phi_1} = \ket{\phi_1}$ (for a detailed discussion on the quantum circuit implementation of $\ket{\phi_0}$ and $\ket{\phi_1}$ refer to Sec.~\ref{sec:init_state} of the \textit{Appendix}). At this point, full translational invariance can be recovered using Linear Combination of Unitaries (LCU)~\cite{Childs2012}, as described below.

\subsection{Symmetrization by Linear Combination of Unitaries (LCU)}\label{sec:LCU}
Given a variational state $\ket{\Psi_{\theta}}$ satisfying $\mathcal{\hat{T}}^2\ket{\Psi_{\theta}}=\ket{\Psi_{\theta}}$, we define the following linear combination to represent normalized, translationally invariant states with momentum $k=0$ or $k=\pi$ as:
\begin{equation}\label{eq:LCU}
    \ket{\Psi_{\theta}^{k}} = \frac{\ket{\Psi_\theta} + e^{ik}\mathcal{\hat{T}}\ket{\Psi_\theta}}{||\ket{\Psi_\theta} +e^{ik} \mathcal{\hat{T}}\ket{\Psi_\theta}||} \ .
\end{equation}
Starting from the definition of $\ket{\Psi_{\theta}^{k}}$ it is easy to show that $\mathcal{\hat{T}}\ket{\Psi_{\theta}^{k=0}}=\ket{\Psi_{\theta}^{k=0}}$ and $\mathcal{\hat{T}}\ket{\Psi_{\theta}^{k=\pi}}=-\ket{\Psi_{\theta}^{k=\pi}}$.
In the following discussion, we show how to implement the symmetrized state in Eq.~\eqref{eq:LCU} as a quantum circuit, making use of the LCU technique~\cite{Childs2012}.

The latter method allows the preparation of normalized states in the form $\ket{\Phi} \propto \hat{\Gamma} \ket{\Phi_{0}}$, given a normalized initial state $\ket{\Phi_{0}}$ and a general linear combination of unitaries $\hat{\Gamma}=\sum_{a=0}^{A-1} c_a \hat{U}_a$. Since $\hat{\Gamma}$ is non unitary, the procedure succeeds only with a certain probability, which is related to the normalization constant of $\hat{\Gamma} \ket{\Phi_0}$. To achieve this goal, it is necessary to \textit{unitarize} $\hat{\Gamma}$, a process that involves introducing $\lceil \log_2(A)\rceil$ ancillary qubits, followed by a \textit{projection}.

Focusing on the translational symmetry, given a state invariant under A-site translations, namely $\hat{\mathcal{T}}^A\ket{\Phi_{0}}=\ket{\Phi_{0}}$, symmetrizing it necessitates $\hat{\Gamma} = \sum_{a=0}^{A-1} e^{ika} \ \hat{\mathcal{T}}^a$, where $k=2\pi n/A$ with $n=0, \dots , A-1$ (see Sec.~\ref{sec:LCU_comp_cost_general} of \textit{Appendix}). This procedure requires a logarithmic number of ancillary qubits $\lceil \log_2(A)\rceil$ and $O(A \cdot N)$ quantum operations independently of the state $\ket{\Phi_{0}}$. Moreover, as this procedure succeeds only with a fixed probability, it may involve a sampling overhead. Nevertheless, it can be shown (see Sec.~\ref{sec:LCU_comp_cost_general} of \textit{Appendix}) that the \emph{average} probability of success across various initializations of the variational parameters is approximately $1/A$, with corrections of order $O(2^{A-N})$. This suggests that, at the beginning of the optimization, the sampling overhead is expected to scale linearly with $A$, but roughly independently of the system size $N$ (assuming $N \gg A$). 

In our case, we implement Eq.~\eqref{eq:LCU} by choosing $A=2$, $\ket{\Phi_0} = \hat{U}_\theta \ket{\phi}$ and $\hat{\Gamma} = \hat{\mathbb{1}} + \hat{\mathcal{T}}$. Consequently, only one ancilla is required, regardless of the system size $N$. Furthermore, the average success probability approximate $1/2$, regardless of $N$, and as a consequence the computational cost of a posteriori symmetrization is kept constant without recurring to \emph{oblivious amplitude amplification}~\cite{Berry2014} or related techniques (see Sec.~\ref{sec:LCU_comp_cost_general} of the \textit{Appendix} for a numerical test). The corresponding circuit is schematically illustrated in Fig.~\ref{fig:circuits}c.
First, we prepare the ancilla in the $\ket{+}$ state. Then, we apply a \emph{controlled} version of $\hat{\mathcal{T}}$. Finally, the projection is performed by measuring the ancilla in the $\{ \ket{+}, \ket{-}\}$ basis. Depending on the measurement outcome, the computational register is prepared in either $\ket{\Psi_\theta^{k=0}}$ or $\ket{\Psi_\theta^{k=\pi}}$ states. Thus, by post-selecting the appropriate measurement results from the ancillary qubit, both momenta ($k=0$ or $k=\pi$) can be obtained using just one circuit\footnote{In Sec.~\ref{sec:LCU_comp_cost} of the \textit{Appendix} we discuss both the computational cost of LCU to restore translational symmetry on a two-site translationally invariant state and to enforce a generic discrete symmetry.}. Additionally, we mention that other symmetrization approaches, based on classical post-processing, are possible but may require a higher number of circuit evaluations~\cite{Seki2020}. 

While effective in the ideal case, \textit{a posteriori} symmetrization procedures (like those discussed in this Section) are notably sensitive to errors arising from noise~\cite{Carrera_Vazquez_2023, Chakraborty2023}. Furthermore, since the circuit depth scales \emph{linearly} with the system size $N$, scaling our approach in the presence of noise may be challenging. Nevertheless, energy level crossings are expected to be effective at relatively moderate system sizes. Therefore, by combining small sized circuits with error mitigation techniques, this approach could still effectively detect the phase transition point, even when using imperfect devices. For this reason, in the following, we investigate the impact of small perturbations on the quantum circuit (see Sec.~\ref{sec:noise} and Sec.~\ref{sec:role_of_noise}). 

\section{Noise and mitigation}\label{sec:noise}
Quantum devices are susceptible to noise, leading to errors during quantum computation. Broadly speaking these errors can be divided into two classes: coherent and incoherent. Coherent errors arise from the miscalibration of quantum gates, resulting in slight shifts in gate rotation angles~\cite{krantz2019quantum}. In contrast, incoherent errors arise from interactions with the environment~\cite{krantz2019quantum,breuer2002theory}. 

\subsection{Noise model}
Accurately modeling incoherent errors is challenging~\cite{PhysRevA.104.062432,dibartolomeo2023novel,vischi2023simulating}. Here, we define a simple noise model that captures only local incoherent errors, allowing to examine how small perturbations affect the symmetries of the variational state. This noise model is implemented as quantum channels $\mathcal{E}_{\tau}$, which generally depend on the gate time $\tau$. They are used to approximate the dominant local errors typically occurring on real devices during the gate execution. The most common models include depolarization $\mathcal{E}_{\tau}^{D}$ and thermal relaxation $\mathcal{E}_{\tau}^{R}$, which are applied after each gate in the quantum circuit~\cite{nielsen2000quantum,benenti2019principles}. Both maps induce single-qubit decoherence. However, the fixed points of the two maps differ, thus introducing competing effects. On the one hand, single-qubit depolarization tends to bring the state towards $\mathbb{1}/2$, namely the maximally mixed one. On the other hand, thermal relaxation tends to bring the state to $\ket{0}\bra{0}$ (see Sec.~\ref{sec:kraus_map} of the \textit{Appendix} for a formal definition and a discussion of the noise channels). Here, we choose to combine the two as $\mathcal{E}_{\tau} = \mathcal{E}^{D}_{\tau} \circ \mathcal{E}^{R}_{\tau}$. Furthermore, we neglect cross talks and correlated noises~\cite{Sarovar2020detectingcrosstalk} implying that the total channel associated to $m$-qubit gates is $\mathcal{E}_{\tau}^{\otimes{m}}$, i.e., the tensor product of the single-qubit one.

Together with errors arising from the gate execution, readout errors are also present during the measurement procedure at the end of computation~\cite{naghiloo2019introduction}. However, we neglect this error source, since highly effective techniques such as T-REX~\cite{van2022model}, capable of mitigating these errors, are already accessible on current quantum devices~\cite{ibm_quantum_exp}. Moreover, since such quantum devices can only implement a specific set of universal gates (see Sec.~\ref{sec:kraus_map} of the \textit{Appendix} ), to conduct numerical simulations that are more faithful to the real hardware, we transpiled our algorithm accordingly.

\subsection{Zero Noise Extrapolation (ZNE)}
In the noise model outlined previously, the quantum channels $\mathcal{E}_{\tau}$ depend on the gate time $\tau$ and the error probabilities, associated to each channel, increases with $\tau$ (see Sec.~\ref{sec:kraus_map} of the \textit{Appendix}). This parameter can be theoretically adjusted to control the noise level, a manipulation that can also be realized experimentally through various techniques~\cite{PhysRevLett.119.180509,unitaryfolding,Giurgica_Tiron_2020}. This allows to perform Zero Noise Extrapolation (ZNE)~\cite{PhysRevLett.119.180509}, an error mitigation strategy suitable for expectation value estimations. In this approach, the expectation value is computed with increasing noise levels (i.e., the gate time $\tau$), in order to extrapolate the ideal result in the zero-noise limit (i.e., $\tau \rightarrow 0$). Generally, the introduction of noise explicitly breaks the symmetries of the Ansatz~\cite{Tüysüz2024}, implying possibly wrong estimations of the low-lying excited energies. To counteract this effect we modify the VQE cost function with a penalty term which favours variational states lying in the correct symmetry sector~\cite{Lyu_2023}. By exploiting the optimizer to mitigate the noise-induced effects on the symmetries of the Ansatz, in combination with ZNE, we accurately recover the zero-noise limit.

\begin{figure}[t]
\centering
\includegraphics[width=\columnwidth]{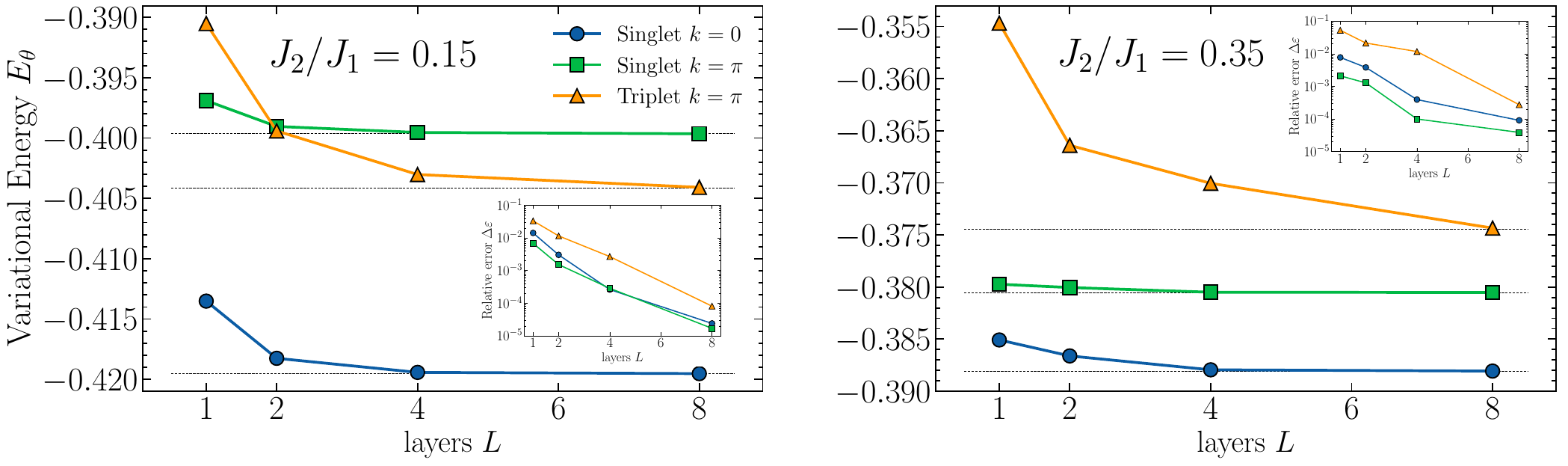}
  \caption{\label{fig:layers} The variational energy for a cluster of $N=16$ sites as a function of the number of layer $L = 1, \dots, N/2$ for $J_2/J_1 = 0.15$ (left panel) and $J_2/J_1=0.35$ (right panel). The exact energies are also reported as dotted lines in both panels. The corresponding relative error $\Delta \varepsilon = |E_{\theta} - E_{\text{ex}}|/E_{\text{ex}}$ with respect to the exact energies is reported as insets as a function of $L$.}
\end{figure}

\section{Noiseless Numerical Results}\label{sec:noiseless}
In the following we examine the results obtained in the noiseless scenario by performing numerical state-vector simulations (PennyLane's \texttt{lightning.qubit} backend~\cite{pennylane2022}). Specifically, we investigate how the accuracy of the variational state can be systematically enhanced by increasing the number of layers $L$ in the circuit across the different symmetry sectors. 

We focus on a cluster of $N=16$ qubits for two frustration ratios: $J_2/J_1 = 0.15$ (gapless phase) and $J_2/J_1 = 0.35$ (gapped phase). In Fig.~\ref{fig:layers} we show the dependence of the variational energies with respect to the number of layers $L$ of the circuit. The optimizations are carried out fixing the quantum numbers of total spin $S$ (singlet or triplet) and momentum ($k=0, \pi$). For both frustration ratios, the ground state is a singlet ($S=0$) with $k=0$.
The first excited state before the phase transition ($J_2/J_1=0.15$, left panel) is a triplet state ($S=1$) with momentum $k=\pi$. Then, after the transition ($J_2/J_1=0.35$, right panel), it becomes a singlet state ($S=0$) with momentum $k=\pi$. The situation is reversed for the second excited state.

For both frustration ratios, the relative error of the states compared to the exact ones~\cite{lanczos1950iterative, sandvik2010} is of order $\Delta \varepsilon \approx 0.01\%$ for a number of layers $L = N/2$~\cite{ho2019} (see insets in Fig.~\ref{fig:layers}). We observe that the convergence of the triplet state is slower compared to the singlet case (see Fig.~\ref{fig:layers}). This behavior can be attributed to the larger symmetry sector of the triplet state ($S=1$), which includes $S^z = -1, 0, 1$, in contrast to the singlet excited state $S=0$, which only includes $S^z = 0$. Consequently, more layers are required to adequately span the entire symmetry subspace~\cite{Lyu_2023}.
Nevertheless, by reaching $L = N/2$, the accuracy is of the same order of magnitude for both cases. Additionally, we emphasize that using a highly symmetric Ansatz, as we do here, generally reduces the subspace that needs to be explored during optimization. This approach helps mitigate trainability challenges in variational circuits, such as barren plateaus, where the variance of the cost function becomes exponentially small increasing the system size~\cite{Ragone24,Fontana24}. Furthermore, we employ a Gaussian initialization of the parameters, with variance scaling as $\sim 1/L$, which was shown to be effective to mitigate barren plateaus in the general setting~\cite{Zhang2022}.
The accurate results obtained before and after the transition suggest that the variational state accurately captures the transition point through the crossing of excited states.

\begin{figure}[t]
    \center
     \includegraphics[width=\columnwidth]{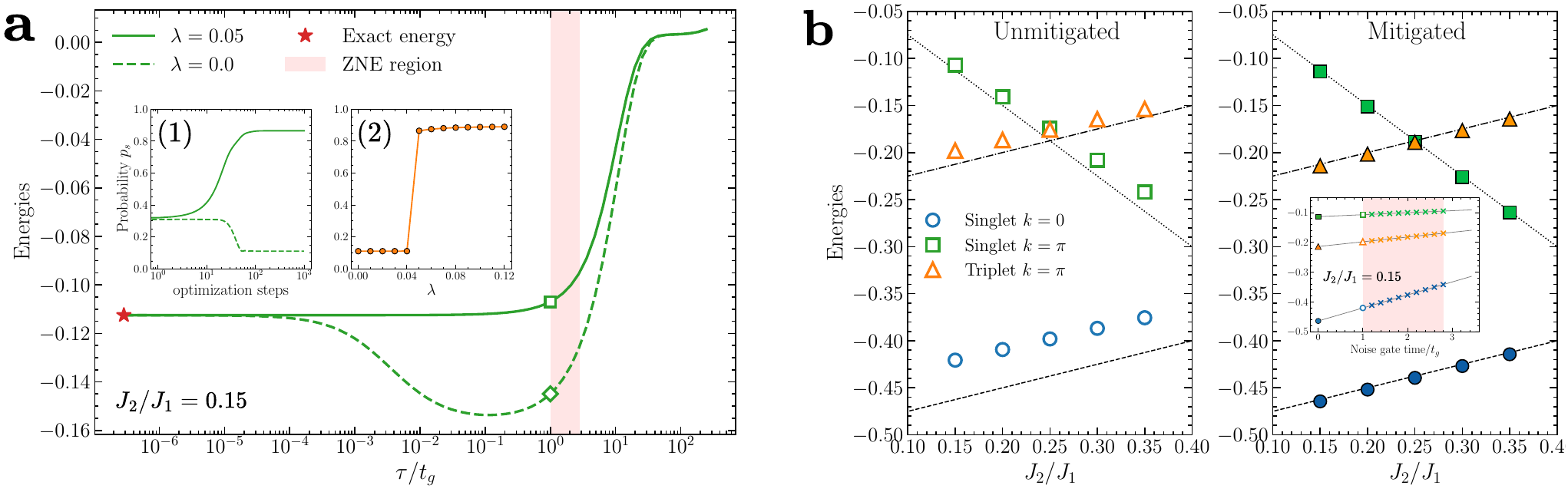}
  \caption{\label{fig:noise_results} \textbf{a)} Variational energies for $S=0$ and $k=\pi$ on a $N=4$ site cluster at $J_1/J_2=0.15$, plotted as a function of $\tau/t_g$, with $\lambda=0.0$ (dashed green curve) and $\lambda=0.05$ (solid green curve). The empty green square (diamond) indicates the energy at $\tau/t_g=1$ for $\lambda=0.05$ ($\lambda=0.0$). The shaded red region marks the \textit{ZNE region} spanning $\tau/t_g \in [1,3]$. The exact energy value is also shown for comparison (red star).
  The inset (1) displays the evolution of the probability $p_\text{s}$ over optimization steps for the gate time $\tau=t_g$, with $\lambda=0.0$ (dashed green curve) and $\lambda=0.05$ (solid green curve). The inset (2) shows the values of the probability $p_\text{s}$, at the end of the optimization, as a function of $\lambda$. \textbf{b)} Variational energies for the different symmetry sectors in the interval $J_2/J_1 \in [0.15,0.35]$. Unmitigated results are obtained at $\tau=t_g$ (left panel), while the mitigated energies obtained via ZNE are shown in the right panel. The inset shows ZNE extrapolations performed within the \textit{ZNE region} ($\tau/t_g \in [1,3]$) for $J_1/J_2 = 0.15$.}
\end{figure}

\section{Noisy Numerical Results}
\label{sec:role_of_noise}
The introduction of noise in the circuit modifies the performance of the Ansatz. In this section we investigate the impact of the noise model (see Sec.~\ref{sec:noise}) on the symmetries of the variational state. We show that it is possible to mitigate errors through ZNE combined with the introduction of a suitable penalty term in the cost function. Here, we focus on a system of $N=4$ sites using a circuit with $L=1$ layer. Indeed, for such a small cluster one layer is sufficient to get accurate estimations in the noiseless limit. Using as reference current IBM superconducting devices~\cite{ibm_quantum_exp}, the single-qubit gate time is $t_g \approx 3.5 \times 10^{-8}$ s. For this reason, variational optimizations are performed setting $\tau = t_g$ (see Sec.~\ref{sec:kraus_map}). Regarding mitigation, we dub \textit{ZNE region} the gate-time interval attainable to perform ZNE on current quantum hardware (i.e. from $\tau=t_g$ up to $\tau\approx 3t_g$). All numerical simulations are performed by employing a density matrix simulator (PennyLane's \texttt{default.mixed} backend~\cite{pennylane2022}).

\subsection{Breaking of Equivariance due to Noise}
In Fig.~\ref{fig:noise_results}a we show the results at $J_2/J_1=0.15$ for the singlet excited state ($S = 0$) at $k = \pi$.
We start describing what happens when changing continuously the gate time $\tau$ in the variational circuit. 
First, we optimize the variational state at $\tau=t_g$ 
(empty green diamond). Then, fixing the optimal parameters, the noise level is changed shifting $\tau$ in the interval $\tau/t_g \in [10^{-7},10^{2}]$ (dashed green curve). We point out that for small values of the gate time ($\tau/t_g \sim 10^{-7}$) the variational result approaches the exact energy (red star). 

However, in the \textit{ZNE region} (shaded interval) the variational energy is lower with respect to the exact one in the selected symmetry sector\footnote{We remark that the variational energy is consistently higher than the ground state energy. However, if the symmetries of the variational state are not preserved, its energy may be lower than that of the excited state we are approximating.}.
As a result, due to the non-monotonic behaviour of the noisy energy curve, performing the ZNE in the \textit{ZNE region} gets a off-target result, committing an error of $45 \%$ with respect to the exact energy. This suggests that the variational state has no definite momentum, implying a breaking in the equivariance of the circuit due to the presence of noise~\cite{Tüysüz2024}.

To better understand this behaviour, we measure, during the energy optimization at $\tau = t_g$, the probability $p_{\text{s}}(\theta) = (1+e^{ik}\bra{\Psi_{\theta}}\hat{\mathcal{T}}\ket{\Psi_{\theta}})/2$ (see Sec.~\ref{sec:LCU_comp_cost_usecase} of the \textit{Appendix} for a derivation), which quantifies the success in performing the LCU symmetrization (green dashed curve in the inset (1) of Fig.~\ref{fig:noise_results}a).  This probability decreases during the optimization and at the end is quite low ($p_{\text{s}} \approx 15 \%$). 
We point out that in a noiseless scenario, the definition of a state with a definite momentum is achieved irrespective of the value of $p_s$. However, when noise is present, the probability of success $p_s$ in executing the LCU symmetrization becomes relevant to the effective generation of translationally invariant states. In general for both scenarios, low values of $p_s$ imply that the one-site translations are primarly restored by LCU, indicating that components in the wrong symmetry sectors are relevant before the LCU application. Conversely, when $p_s$ is high, most of the symmetrization is effectively performed by optimizing the parameters within the variational circuit $\hat{U}_\theta$, relegating a minor role to LCU. Since the latter is the most susceptible component to noise in the quantum circuit~\cite{Carrera_Vazquez_2023, Chakraborty2023}, its contribution to equivariance loss is expected to be predominant. Consequently, we identify the decay of $p_s$ during training as the main indicator of noise-induced equivariance breaking within the circuit.

\subsection{Noise mitigation and symmetry restoration}

In this section, we devise a strategy to mitigate the effect of noise on the symmetries of the Ansatz. As discussed in the previous section, in order to reduce the role of the LCU in the construction of a translationally invariant state we aim at increasing the probability $p_s$. This can be achieved by adding a penalty term in the loss function
\begin{equation}\label{eq:penalty}
    P(\lambda,\theta) = \lambda \left[ 1 - p_{\text{s}}(\theta)\right]^2 \ .
\end{equation}
Here, $\lambda$ is an hyperparameter which controls the intensity of the penalty term. Performing energy minimizations at different values of $\lambda$, fixing the gate time $\tau=t_g$, allows to determine its optimal value.
In the inset (2) of Fig.~\ref{fig:noise_results}a we show a sharp transition at $\lambda=0.05$ from a regime of low ($p_{\text{s}}\approx 15 \%$) to high ($p_{\text{s}} \approx 90 \%$) probability measured at the end of the training. 
In addition, in the inset (1) we show how the behaviour of the probability $p_{\text{s}}$, during the energy optimization, is modified by introducing a penalty term (solid green curve). As a result, by setting $\lambda=0.05$, the energy of the optimized state (empty green square in Fig.~\ref{fig:noise_results}a) results in a reliable approximation of the energy of the singlet excited state at $k=\pi$ (with a relative error $\Delta \varepsilon \approx 5 \%$). 

At this stage, maintaining the optimal variational parameters obtained at $\tau=t_g$, the noise level is changed shifting $\tau$ in the interval $\tau/t_g \in [10^{-7},10^{2}]$ (green solid curve in Fig.~\ref{fig:noise_results}a). Here, a monotonic behaviour emerges, facilitating the implementation of Zero Noise Extrapolation within the \textit{ZNE region} (see below).

\subsection{Mitigated energy level crossing}
Finally, in Fig.~\ref{fig:noise_results}b we estimate the variational energies, at gate time $\tau =t_g$, for the different symmetry sectors varying the values of the frustration ratio in the interval $J_2/J_1 \in [0.15, 0.35]$\footnote{In particular, an appropriate value of $\lambda$ is chosen for each simulation.}. As shown in the left panel, the variational energies are shifted with respect to the exact ones (marked by dashed lines) due to noise. The mitigated results with ZNE technique performed in the \textit{ZNE region} are depicted in the right panel.

In the inset, for $J_2/J_1=0.15$, the empty points represent the noisy energies (also depicted on the left panel), while the crosses denote the expectation values obtained with increasing gate time. The filled points indicate the extrapolated values in the zero-noise limit (also displayed in the right panel) after fitting the data by linear regression.

We point out that even in the unmitigated scenario (left panel), the crossing point is adequately captured, despite slight energy shifts. However, through mitigation, we not only identify the energy level crossing accurately but also approximate the exact energies with an $\Delta \varepsilon \approx 1\%$ error (right panel).

\section{Conclusions and Outlook}
We have introduced an Ansatz inspired by HVA to investigate the excited states of an interacting spin model on a lattice. Specifically, we have discussed the possibility to carry out optimizations in specific symmetry sectors by fixing the quantum numbers in the variational state. While HVA states are not yet competitive with classical methods, such as Tensor Networks and NQS, for the approximation of the low-energy states of quantum spin models, this work aims to emphasize that the total-spin symmetry, generally challenging to incorporate in classical approaches~\cite{vieijra2020,vieijra2021,singh2012,schmoll2020}, can be more straightforwardly implemented in quantum circuit-based Ans\"atze. Although highly accurate results can often be achieved with classical states even without enforcing exact total-spin conservation, we believe that having a simple method to construct $SU(2)$-symmetric states presents an interesting application of HVA quantum circuits.
It is worth noting that, by employing level spectroscopy, the quantum phase transition point in the one-dimensional $J_1$-$J_2$ Heisenberg model is estimated at $J_2/J_1 = 0.25$ on a cluster of $N=4$ sites (see Fig.~\ref{fig:noise_results}b) with an error of $3.6\%$ compared to the thermodynamic limit result $(J_2/J_1)_{\text{c}} = 0.24117(6)$. Given the negligible size effects of this technique, it can be properly generalized to study other models, specifically the application of the same approach to systems with long-range interactions or two-dimensional systems, where the presence of phase transitions is still under debate, is the focus of future investigations. 

Furthermore, we discussed how noise, which explicitly breaks state symmetries, can be mitigated by standard techniques such as ZNE, with the addition of an appropriate penalty term which helps in finding symmetric solutions. From this perspective, the implementation of this Ansatz on current quantum devices represents the next step in verifying whether the mitigation techniques employed in simulations are still effective. Moreover, this implementation will involve assessing also the impact of finite samples on the estimation of the expectation values~\cite{astrakhantsev2023}.

\section*{Acknowledgments}
We thank F. Becca, A. Bassi, M. Grossi, A. Sandvik, M. Imada and R. Kaneko for useful discussions. G.C., G.D.B. and M.V. acknowledge the financial support from University of Trieste and INFN.
The numerical simulations have been
performed within the PennyLane library~\cite{pennylane2022}.

\clearpage
\appendix

\section*{Appendix}
\section{Initial states}\label{sec:init_state}

\subsection{Singlet state $\ket{\phi_0}$}
\begin{figure}[b]
    \centering
    \includegraphics[width=\columnwidth]{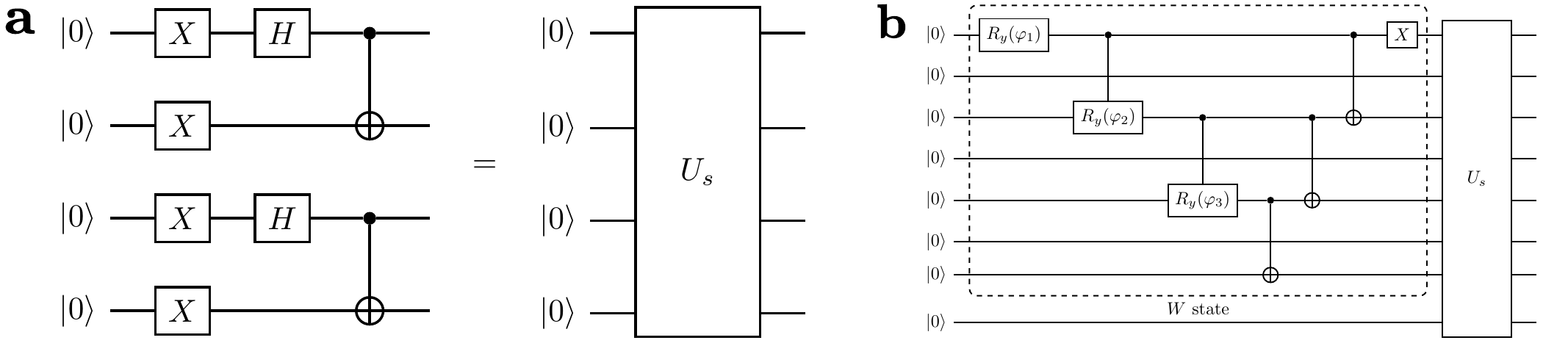}
    \caption{\textbf{a)} Circuit $\hat{U}_s$ for preparing the $\ket{\phi_0}$ state on 4 qubits. \textbf{b)} Circuit for preparing the $\ket{\phi_1}$ state on 8 qubits.}
    \label{fig:singlets_triplets}
\end{figure}

The initial state representing factorized singlet pairs, as given in Eq.~\eqref{eq:singlet_state}, is easily prepared on a quantum computer. Simultaneously for each singlet pair $2r-1,2r$, we prepare the state $\ket{1,+}$ using Pauli $X$ and Hadamard $H$ gates, followed by a CNOT gate (refer to Fig.~\ref{fig:singlets_triplets}a). In general, we dub $\hat{U}_s$ the unitary transformation responsible for generating the product of singlet state on $N$ qubits.

\subsection{Triplet state $\ket{\phi_1}$}
The triplet states in Eq.~\eqref{eq:triplet_state_tilde} are not two-site translation invariant. To get the invariant state $\ket{\phi_1}$ in Eq.~\eqref{eq:triplet_state} we use a particular class of entangled states called $W$ states \cite{dur2000}. The $W$ state for $n$ qubits is the superposition of all the possible $n$-qubit states with just a single qubit in the state $\ket{1}$
\begin{equation}
    \ket{W^{(n)}} = \frac{1}{\sqrt{n}}\big(\ket{10\dots0} + \ket{01\dots0}+\dots + \ket{00\dots1}\big).
\end{equation}
As an example for the preparation of $\ket{\phi_1}$, we consider the case of $N=4$ qubits. We set qubits 2 and 4 in the state $\ket{00}$ and qubits 1 and 3 in the $W$ state, that reads
\begin{equation}
    \ket{W^{(2)}}_{13} = \frac{1}{\sqrt{2}}\big(\ket{10}_{13} + \ket{01}_{13}\big).
\end{equation}
Then, the total state is expressed as
\begin{equation}
\label{eq:W_state_4_qubits}
\ket{W^{(2)}}_{13}\ket{00}_{24}= \frac{1}{\sqrt{2}}\big(\ket{1000}+\ket{0010}\big) \ ,
\end{equation}
where in the right-hand side of Eq.~\eqref{eq:W_state_4_qubits} the states of single qubits are indexed in ascending order.
By applying the unitary $\hat{U}_{s}$ to the state in Eq.~\eqref{eq:W_state_4_qubits}  we get
\begin{equation}
\label{eq:triplet_4_qubits}
\frac{1}{\sqrt{2}}\bigl(\ket{t}_{12}\ket{s}_{34} + \ket{s}_{12}\ket{t}_{34}\bigr)= \frac{1}{\sqrt{2}}\Bigl(\ket{\tilde{\phi}_1^1} + \ket{\tilde{\phi}_1^2}\Bigr),
\end{equation}
where the right-hand side of Eq.~\eqref{eq:triplet_4_qubits} is $\ket{\phi_1}$ for $N=4$ qubits.
Thus in general, to get $\ket{\phi_1}$ all odd indexed qubits are prepared in the $W$ state and this transformation is followed by the singlet preparation $\hat{U}_s$ on all qubits. An example of the resulting circuit for eight qubits is shown in Fig.~\ref{fig:singlets_triplets}b, where the preparation of the $W$ state is performed efficiently with the techniques described in Ref.~\cite{mcclung2020constructions}. The angles $\varphi_{i}$ in Fig.~\ref{fig:singlets_triplets}b are defined as $\varphi_{i} = 2\arccos({1/\sqrt{n-i+1}})$ where $i$ is the index of the qubit.

\section{Alternative choice of $\hat{H}_1$ and $\hat{H}_2$}\label{sec:init_ham}
As detailed in Sec.~\ref{sec:vqe}, we choose the Hamiltonians $\hat{H}_1$ and $\hat{H}_2$ to define the variational circuit in Eq.~\eqref{eq:unitary} as $\hat{H}_{\text{odd}}$ and $\hat{H}_{\text{even}}$, respectively. Together, they constitute the nearest-neighbor contribution in the $J_1$-$J_2$ Heisenberg model [refer to Eq.~\eqref{eq:J1J2_ham}]. However, an alternative choice for $\hat{H}_1$ and $\hat{H}_2$ is feasible, incorporating contributions from the next-nearest neighbor term $\sum_{r=1}^N \hat{\boldsymbol{S}}_{r}\cdot\hat{\boldsymbol{S}}_{r+2}$.
For $J_2/J_1=0.5$, the Hamiltonian in Eq.~\eqref{eq:J1J2_ham} corresponds to the exactly solvable Majumdar–Ghosh model~\cite{Majumdar1969}, denoted here as $\hat{H}_{\text{MG}}$. In this model, the ground state is represented by a product of singlet pairs, and the first excited state can be obtained by replacing one singlet pair with a triplet one, analogously to the Hamiltonian $\hat{H}_{\text{even}}$ (see Sec~\ref{sec:vqe}). Consequently, the $J_1$-$J_2$ Hamiltonian can be expressed as
\begin{equation}
    \hat{H}_{J_1\text{-}J_2} = \hat{H}_{MG} + \left(J_2 - \frac{J_1}{2}\right) \sum_{r=1}^N \hat{\boldsymbol{S}}_{r}\cdot\hat{\boldsymbol{S}}_{r+2} \ .
\end{equation}
Accordingly, we can define a variational quantum circuit in the form of Eq.~\eqref{eq:unitary} setting $\hat{H}_2 =\hat{H}_{\text{MG}}$ and $\hat{H}_1 = \sum_{r=1}^N \hat{\boldsymbol{S}}_{r}\cdot\hat{\boldsymbol{S}}_{r+2}$. 

\begin{figure}[t]
    \center
     \includegraphics[width=\columnwidth]{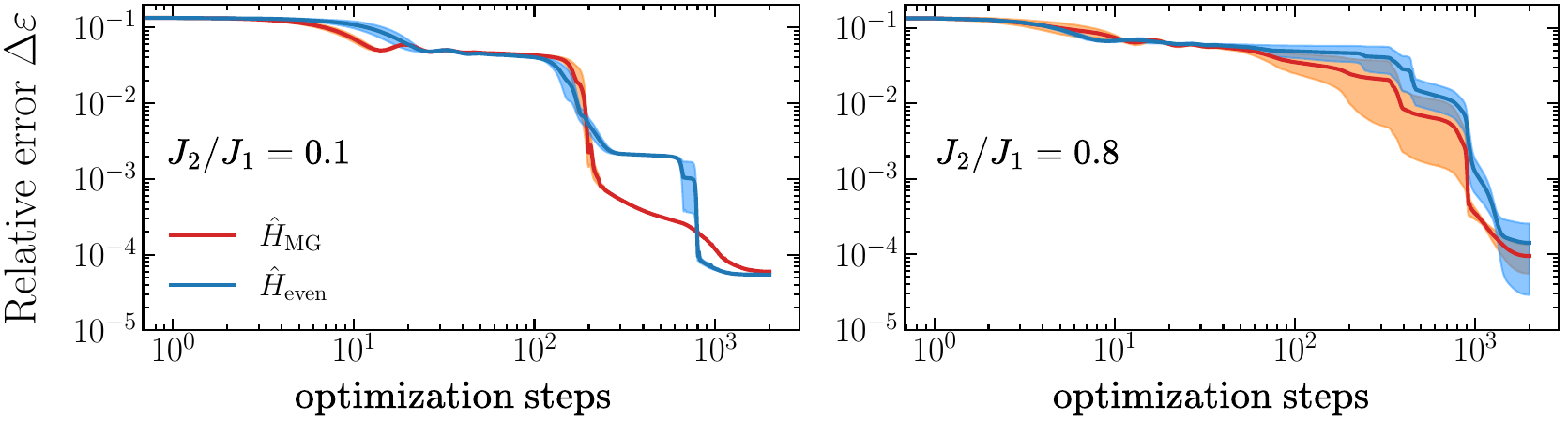}
  \caption{\label{fig:MG} Comparison of the accuracy of variational energies, relative to exact values, for the states defined by $\hat{H}_{\text{even}}$ and $\hat{H}_{\text{MG}}$ as a function of the optimization steps. Optimizations are performed on a cluster of $N=10$ sites for $J_2/J_1=0.1$ (left panel) and $J_2/J_1=0.8$ (right panel). The mean behaviours over multiple optimizations are represented by solid red and blue lines for the variational states related to $\hat{H}_{\text{MG}}$ and $\hat{H}_{\text{even}}$, respectively. Orange and light-blue intervals indicate the corresponding standard deviations.}
\end{figure}

In Fig.~\ref{fig:MG}, we illustrate a comparison of the accuracy of the variational states with the two different choices of $\hat{H}_2$ and $\hat{H}_1$. Specifically, we plot the relative error of the ground state energy for a cluster of $N=10$ sites as a function of the optimization steps. The left panel represents the results for a small value of the frustration ratio $J_2/J_1=0.1$, while the right panel corresponds to a large value $J_2/J_1 = 0.8$. For each frustration ratio, multiple optimizations are performed using $L=N/2$ layers for $J_2/J_1=0.1$ and $L = N$ layers for $J_2/J_1=0.8$.
The relative error of the two variational states is of the same order of magnitude ($\Delta \varepsilon \approx 0.01 \%$) at the end of the optimizations.

Given that the variational state defined using $\hat{H}_{\text{MG}}$ includes gates that directly entangle next-nearest neighbors qubits, we might expect that it is more accurate for large values of the frustration ratio ($J_2/J_1 > 0.5$) compared to the one defined in Sec.~\ref{sec:vqe}.
However, even for $J_2/J_1=0.8$, the two variational states yield similar results. This could be attributed to the fact that increasing the number of layers $L$ in the circuit also the state defined through $\hat{H}_\text{even}$ entangles (indirectly) next-nearest neighbor sites. As a result, the two unitaries produce comparable variational results.

Moreover, by parameter sharing among different qubits the circuit corresponding to the Ansatz with $\hat{H}_\text{MG}$ is translational invariant of four lattice sites. Consequently, constructing a translational invariant state with defined momentum $k=0$ or $k=\pi$ requires a more computationally expensive LCU (see Sec.~\ref{sec:LCU_comp_cost_general} of the \textit{Appendix}). Therefore, given the similar accuracy of the two states, the numerical calculations in this work were carried out using the the circuit defined in the main text.

\section{Computational costs of \emph{a posteriori} symmetrization}
\label{sec:LCU_comp_cost}

In this Section, we consider the asymptotic increase in quantum resources due to symmetrization as a function of the number $N$ of qubits. This analysis is conducted in terms of:
\begin{enumerate}
    \item \textit{Gates overhead}. We determine the scaling of quantum operations not included in the state preparation $\hat{U}_\theta$.
    \item \textit{Samples overhead}. We analyze the success probability $p_{\text{s}}$ of the algorithm. Indeed, achieving (on average) a number of accepted samples of $n_{s}$ requires performing a \emph{larger} number of experiments $n_{{\text{sym}}}$ such that $p_{\text{s}}\,n_{{\text{sym}}} = n_{s}$. This gives a sampling overhead inversely proportional to $p_{\text{s}}$.
\end{enumerate}

Here, we explicitly compute these overheads, both in our use case (Sec.~\ref{sec:LCU_comp_cost_usecase}) and in general (Sec.~\ref{sec:LCU_comp_cost_general}).

\subsection{Restoring full translational invariance given two-site invariance}
\label{sec:LCU_comp_cost_usecase}

As show in Sec.~\ref{sec:LCU}, LCU can be used to restore full translational symmetry in a variational state $\ket{\Psi_\theta}$ by applying the circuit summarized in Fig.~\ref{fig:circuits}c. In particular, this holds provided that the initial state is already invariant under two-site translations, i.e.,  $\hat{\mathcal{T}}^2\ket{\Psi_\theta} = \ket{\Psi_\theta}$.\\

1. \textit{Gates overhead}. Additional operations include preparation of the $\{\ket{+}, \ket{-}\}$ states in the ancilla [$O(1)$ operations] and implementation of $\hat{\mathcal{T}}$, which requires $N-1$ SWAP gates [$O(N)$ operations]. Note that in the controlled version, instead of SWAPs, one needs to implement Fredkin gates (controlled SWAPs), which although more expensive, still require $O(N)$ elementary operations overall, yielding a total cost of $O(N)$ gates.\\

2. \textit{Samples overhead}. Consider the preparation of variational states of momentum $k=0$, namely we post-select samples where the ancilla is found in the state $\ket{+}$.
In this case, $p_{\text{s},k=0}$ can be computed using standard LCU theory
\begin{equation}
\label{eq:LCU_prob_theory}
    p_{\text{s}} = \frac{\bra{\Phi_0}\hat{\Gamma}^\dagger \hat{\Gamma}\ket{\Phi_0}}{\|\boldsymbol{c}\|^2_1}
\end{equation}
where $\ket{\Phi_0} = \ket{\Psi_\theta}$, $\hat{\Gamma} = \hat{\mathbb{1}} + \hat{\mathcal{T}}$ and $\boldsymbol{c} = (1,1)^T$. After straightforward manipulations, we obtain
\begin{equation}
\label{eq:LCU_prob_re}
    p_{\text{s},k=0}(\theta) = \frac{1}{2} + \frac{\Re{\bra{\Psi_\theta} \hat{\mathcal{T}} \ket{\Psi_\theta}}}{2}.
\end{equation}
Given $\hat{\mathcal{T}} = \hat{\mathcal{T}}^\dagger    \hat{\mathcal{T}}^2$ and the invariance under two-site translations of $\ket{\Psi_\theta}$, it immediately follows that $\bra{\Psi_\theta} \hat{\mathcal{T}} \ket{\Psi_\theta} = \bra{\Psi_\theta} \hat{\mathcal{T}}^\dagger \ket{\Psi_\theta} \in \mathbb{R}$.

Since only two outcomes are possible when measuring a single qubit, it implies ${p_{\text{s},k=\pi} = 1-p_{\text{s},k=0}}$. As a result
\begin{equation}
    p_{\text{s},k}(\theta) = \frac{1}{2} + e^{ik}\frac{\bra{\Psi_\theta}\hat{\mathcal{T}}\ket{\Psi_\theta}}{2} \;\; , \;\; k=0,\pi.
\end{equation}
Note that $p_{\text{s},k}$ depends on the variational parameters, and hence will vary depending on initialization and during the energy minimization.

\subsection{General method for symmetrization by Linear Combination of Unitaries (LCU)}
\label{sec:LCU_comp_cost_general}
\begin{figure}
\centering
\includegraphics[width=\columnwidth]{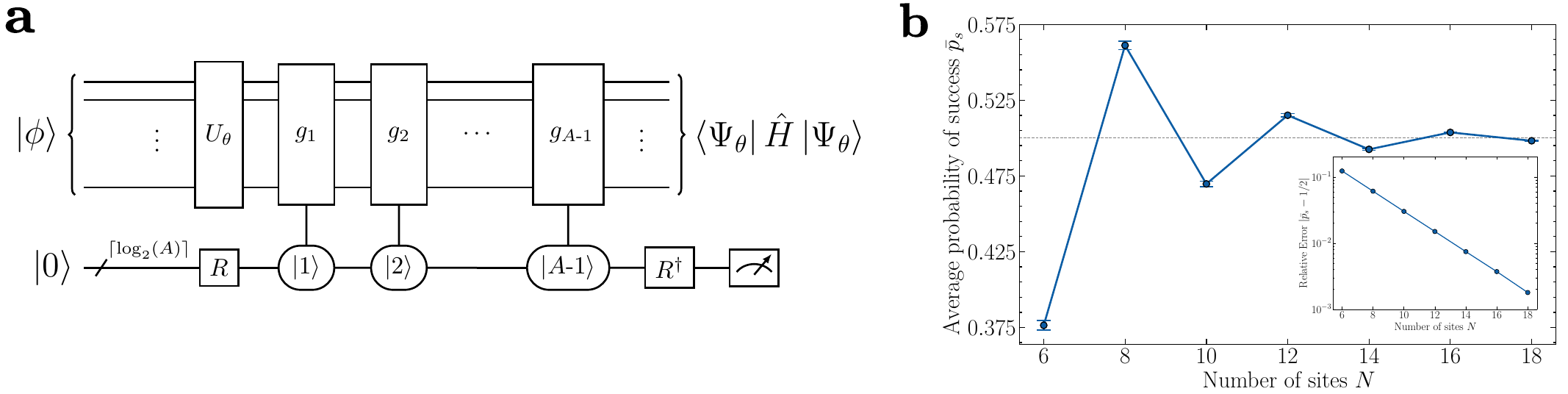}
    \caption{\textbf{a)} General circuit to implement symmetrization by LCU. To control a symmetry operation $\hat{g}_a$ on a specific basis state $\ket{{a}}$ implies the realization of the unitary $\ket{{a}}\bra{{a}}\otimes \hat{g}_a + (\hat{\mathbb{1}}-\ket{{a}}\bra{{a}})\otimes \hat{\mathbb{1}}$. This notation reduces to ordinary controlled gates if $A=2$, i.e. when only one ancillary qubit is required. \textbf{b)} Scaling of the average probability of success $\bar{p}_s$ as a function of the system size $N$ over the initialization distribution for the Ansatz in Sec.~\ref{sec:vqe}, assuming $S = 1$ and $k = \pi$. Inset: Relative error $|\bar{p}_s-1/2|$ as a function of the system size $N$.}
    \label{fig:general_LCU}
\end{figure}

Generalization of the procedure described in Sec.~\ref{sec:LCU} follows directly from the LCU structure. The latter can be used to enforce any discrete, finite group symmetry $G$, with elements ${g}_0,{g}_1, \dots ,{g}_{A-1}$. In general, this is  achieved by applying $\hat{\Gamma}_G = \sum_{a=0}^{A-1} \chi^*_a \hat{g}_a$ where $\chi_a$ is the character and $\hat{g}_a$ is the unitary representation of the of the corresponding group element. The symmetrization procedure is summarized in Fig.~\ref{fig:general_LCU}a, and requires the implementation of $A-1$ controlled operations\footnote{Note that, although $G$ has $A$ elements, the element $g_0$ can always be mapped to the identity $\hat{\mathbb{1}}$, and hence needs not to be implemented.}. The preparation of the ancillary system, in the general case, requires devising a transformation $\hat{R}$ such that 

\begin{equation}
\hat{R} \ket{{0}} = \sum_{a=0}^{A-1} \sqrt{\frac{\chi^*_a}{\sum_{a'}|\chi_{a'}|}}\ket{{a}} \ .
\end{equation}
The procedure is successful only if the ancilla is found in the $\ket{{0}}$ state. \\

1. \textit{Gate overhead}. In the general case, it is not easy to estimate the number of additional operations as a function of $N$ and $A$. In particular, it depends on the number of gates $N_a$ and $N_R$ to implement each controlled $\hat{g}_a$ and $\hat{R}$, respectively.\\

2. \textit{Samples overhead}. A general result can be derived from Eq.~\eqref{eq:LCU_prob_theory}
\begin{equation}
\label{eq:LCU_prob_general}
    p_{\text{s}} = \frac{\sum_a |\chi_a|^2 + 2\sum_{b>a}\Re{\chi_a\chi_b^* \bra{\Psi_\theta}\hat{g}_a^\dagger \hat{g}_b \ket{\Psi_\theta}}}{\left(\sum_a |\chi_a| \right)^2} \ ,
\end{equation}
which directly generalizes Eq.~\eqref{eq:LCU_prob_re}. However, due to the dependence on variational parameters $\theta$, it is not trivial to estimate the scaling of $p_{\text{s}}$. Nevertheless, the average scaling can be computed. In particular, if we assume $\hat{U}_\theta$ forms a unitary 1-design over the initialization probability\footnote{Formally, the distribution of $\hat{U}_{\theta}$ approximates the \emph{uniform} distribution up to the first moment, i.e. $$\mathbb{E}_\theta \left\{ U_\theta \otimes U^*_\theta \right\} = \int_{U(2^N)} d\mu(V) V\otimes V^*$$ where the integration is performed over the uniform (Haar) measure over the unitary group $U(2^N)$.}
\begin{equation}\label{eq:1-design}
    \mathbb{E}_\theta\left\{ \bra{\Psi_\theta}\hat{O} \ket{\Psi_\theta} \right\} = \frac{1}{2^N}\text{Tr}\{\hat{O}\} \ ,
\end{equation}
which holds for any operator $\hat{O}$. This allows to compute the average probability of success $\mathbb{E}_\theta\{p_{\text{s}}\} = \bar{p}_{\text{s}}$. Employing Eq.~\eqref{eq:1-design} we get
\begin{equation}
\label{eq:LCU_prob_general_avg}
    \bar{p}_s = \frac{\sum_a |\chi_a|^2 + 2^{-N}\sum_{b>a}2\Re{\chi_a\chi_b^* \text{Tr}\{\hat{g}_a^\dagger \hat{g}_b}\}}{\left(\sum_a |\chi_a| \right)^2} \ ,
\end{equation}
which now only depends on $N$, $A$ and the representation choice of $G$.\\

Consider as an example $G = \mathbb{Z}_{A}$, namely the \textit{multiplicative group of integers modulo A}. It admits a simple representation given by qubit translations, i.e. $\hat{g}_a = \hat{\mathcal{T}}^a$ and $\chi_a = e^{-ika}$ for $a = 0, \dots ,A-1$ and $k=2\pi n/A$, with $n=0,\dots, A-1$. This symmetry is especially useful when the variational state is already invariant under translations of $A$ sites, and we wish to implement full translational symmetry (e.g., Sec.~\ref{sec:init_ham}). Assuming the implementation of $\hat{R}$ does not scale with $N$, the number of additional operations can be estimated to scale as $O(A \cdot N)$.
Regarding sampling overhead, applying Eq.~\eqref{eq:LCU_prob_general_avg} yields
\begin{equation}
    \bar{p}_{\text{s},k} = \frac{1}{A} + \frac{\sum_{b>a} \cos[ik(a-b)]\text{Tr}\{\hat{\mathcal{T}}^{b-a}\}}{2^{N-1} A^2} \ .
\end{equation}
Furthermore, by noting that $0 \leq \text{Tr} \{\hat{\mathcal{T}}^{b-a}\} \leq 2^{A}$ we can bound the magnitude of the second term:
\begin{equation}\label{eq:LCU_prob_avg_scaling}
    \bar{p}_{\text{s},k}= \frac{1}{A} + O(2^{A-N}) \ .
\end{equation}
This is especially useful in cases where $N \gg A$, since it ensures a sizable initial probability.
Setting $A=2$, we get the use-case discussed in Sec.~\ref{sec:LCU}. While the unitary $1$-design hypothesis is not rigorously met in this case, we numerically show in Fig.~\ref{fig:general_LCU}b that the average probability of success $\bar{p}_s$ approaches $1/2$ exponentially fast as the system size $N$ increases, as predicted by Eq.~\eqref{eq:LCU_prob_avg_scaling}.

\section{Kraus maps}\label{sec:kraus_map}
We report the Kraus maps used for the noisy simulations in the main text.

The Kraus map of single-qubit depolarization reads \cite{benenti2019principles,nielsen2000quantum}
\begin{equation}\label{krausdep}
\mathcal{E}^{D}_{\tau}(\hat{\rho}) =\Bigl(1-\frac{3}{4} p_{\tau}\Bigr) \hat{\rho} + \frac{p_{\tau}}{4} \hat{X}\hat{\rho}\hat{X} + \frac{p_{\tau}}{4} \hat{Y}\hat{\rho}\hat{Y} + \frac{p_{\tau}}{4} \hat{Z}\hat{\rho}\hat{Z}\,,
\end{equation}
where $\hat{\rho}$ is the density matrix of the single-qubit, $\hat{X},\hat{Y},\hat{Z}$ are the Pauli matrices and $p_{\tau}$ quantifies the probability of having a bit flip, a phase flip or a bit-phase flip of the states of the computational basis. We assume a behaviour in time of the form $p_{\tau}= (1 - e^{-\gamma_d \tau})$ for a characteristic time $T_d=1/\gamma_d$ \cite{manzano2020short}.
\begin{figure}
\centering
\includegraphics[width=0.9\columnwidth]{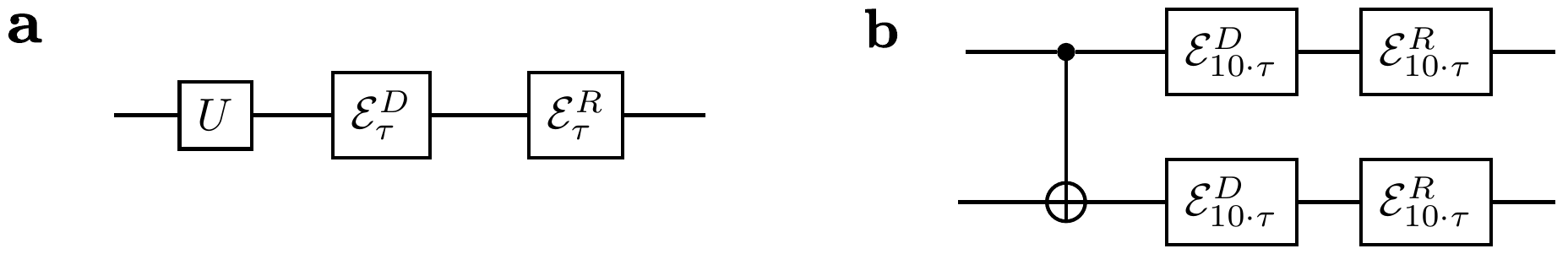}
    \caption{Schematic depiction of the noise model. In panel (a) we show quantum channels associated to a generic single-qubit gate $U$ and in panel (b) those associated to CNOT gates.}
    \label{fig:noise_model}
\end{figure}
The Kraus map of the single-qubit thermal relaxation is given by \cite{benenti2019principles,nielsen2000quantum}
\begin{equation}
\label{relaxKraus}
    \mathcal{E}_{\tau}^{R}(\hat{\rho}) = \hat{K}\hat{\rho}\hat{K} + p^{(1)}_{\tau}\hat{\sigma}^{-} \hat{\rho} \hat{\sigma}^{+} + p^{(z)}_{\tau}\hat{Z}\hat{\rho}\hat{Z}+p^{(1)}_{\tau}\hat{\mathcal{P}}_0\hat{\rho}\hat{\mathcal{P}}_0
\end{equation}
where we define the operators $\hat{K} = \sqrt{1 - p^{(1)}_\tau - p^{(z)}_\tau}\,\, \hat{\mathbb{1}}$, $\hat{\mathcal{P}}_0 = \ket{0}\bra{0}$ and $\hat{\sigma}^{-}$, $\hat{\sigma}^{+}$ are the lowering and raising Pauli operators, respectively. 
Here, $p^{(1)}_{\tau} = (1 - e^{-\tau/T_{1}})$ is the probability of reset to $\ket{0}$ and $T_{1}$ the relaxation time. Additionally, we introduce $p^{(z)}_{\tau} = (1 - p^{(1)}_{\tau}) \cdot p^{(pd)}_{\tau}$, where $p^{(pd)}_{\tau} = (1 - e^{-\tau/T_{pd}})$ is the probability of pure dephasing with $T_{pd} = T_{1}T_{2} / (2T_{1} - T_{2})$ and $T_{2}$ the decoherence time. The time scales $T_{1}$ and $T_{2}$ are related as $T_{2} \le 2T_{1}$~\cite{manzano2020short}.

The circuits used in our simulations are transpiled into the native gate set of IBM devices, i.e., $\{RZ(\alpha), X, \sqrt{X}, \text{CNOT}\}$, where $\alpha \in [-\pi,\pi]$ is a rotation angle. Typically the duration of the execution of CNOT gate is $10$ times larger with respect to the single-qubit gate time. We set the noise time scales as $T_1$ = $T_2 = T_d \approx 10^{-4} \ s$, compatible with the average values of current IBM devices \cite{ibm_quantum_exp} (see Fig.~\ref{fig:noise_model}).

\clearpage
\bibliography{refs}

\providecommand{\newblock}{}
\begin{thebibliography}{10}
\expandafter\ifx\csname url\endcsname\relax
  \def\url#1{{\tt #1}}\fi
\expandafter\ifx\csname urlprefix\endcsname\relax\def\urlprefix{URL }\fi
\providecommand{\eprint}[2][]{\url{#2}}

\bibitem{Sachdev_1999}
Sachdev S 1999 {\em Physics World\/} {\bf 12} 33 \urlprefix\url{https://dx.doi.org/10.1088/2058-7058/12/4/23}

\bibitem{eggert1996}
Eggert S 1996 {\em Phys. Rev. B\/} {\bf 54}(14) R9612--R9615 \urlprefix\url{https://link.aps.org/doi/10.1103/PhysRevB.54.R9612}

\bibitem{sandvik2010}
Sandvik A 2010 {\em AIP Conference Proceedings\/} {\bf 1297} 135--338 \urlprefix\url{https://aip.scitation.org/doi/abs/10.1063/1.3518900}

\bibitem{nomura1995}
Nomura K 1995 {\em Journal of Physics A: Mathematical and General\/} {\bf 28} 5451 \urlprefix\url{https://dx.doi.org/10.1088/0305-4470/28/19/003}

\bibitem{nakamura1999}
Nakamura M 1999 {\em Journal of the Physical Society of Japan\/} {\bf 68} 3123–3126 ISSN 1347-4073 \urlprefix\url{http://dx.doi.org/10.1143/JPSJ.68.3123}

\bibitem{nakamura2000}
Nakamura M 2000 {\em Phys. Rev. B\/} {\bf 61}(24) 16377--16392 \urlprefix\url{https://link.aps.org/doi/10.1103/PhysRevB.61.16377}

\bibitem{sandvik2004}
Sandvik A~W, Balents L and Campbell D~K 2004 {\em Phys. Rev. Lett.\/} {\bf 92}(23) 236401 \urlprefix\url{https://link.aps.org/doi/10.1103/PhysRevLett.92.236401}

\bibitem{sandvik2010b}
Sandvik A~W 2010 {\em Phys. Rev. Lett.\/} {\bf 104}(13) 137204 \urlprefix\url{https://link.aps.org/doi/10.1103/PhysRevLett.104.137204}

\bibitem{wang2018}
Wang L and Sandvik A~W 2018 {\em Phys. Rev. Lett.\/} {\bf 121}(10) 107202 \urlprefix\url{https://link.aps.org/doi/10.1103/PhysRevLett.121.107202}

\bibitem{ferrari2020}
Ferrari F and Becca F 2020 {\em Phys. Rev. B\/} {\bf 102}(1) 014417 \urlprefix\url{https://link.aps.org/doi/10.1103/PhysRevB.102.014417}

\bibitem{nomura2021}
Nomura Y and Imada M 2021 {\em Phys. Rev. X\/} {\bf 11}(3) 031034 \urlprefix\url{https://link.aps.org/doi/10.1103/PhysRevX.11.031034}

\bibitem{franchini2017}
Franchini F 2017 {\em An Introduction to Integrable Techniques for One-Dimensional Quantum Systems\/} (Springer International Publishing) ISBN 9783319484877 \urlprefix\url{http://dx.doi.org/10.1007/978-3-319-48487-7}

\bibitem{white1996}
White S and Affleck I 1996 {\em Phys. Rev. B\/} {\bf 54}(14) 9862--9869 \urlprefix\url{https://link.aps.org/doi/10.1103/PhysRevB.54.9862}

\bibitem{lacroix2011}
Lacroix C, Mendels P and Mila F 2011 {\em Introduction to Frustrated Magnetism: Materials, Experiments, Theory\/} (Springer) ISBN 978-3-642-10588-3

\bibitem{jos2013}
Jos J 2013 {\em 40 Years of Berezinskii-Kosterlitz-Thouless Theory\/} EBL-Schweitzer (World Scientific) ISBN 9789814417648 \urlprefix\url{https://books.google.it/books?id=roO6CgAAQBAJ}

\bibitem{viteritti2022}
Viteritti L~L, Ferrari F and Becca F 2022 {\em SciPost Phys.\/} {\bf 12} 166 \urlprefix\url{https://scipost.org/10.21468/SciPostPhys.12.5.166}

\bibitem{viteritti2023}
Viteritti L~L, Rende R and Becca F 2023 {\em Physical Review Letters\/} {\bf 130} ISSN 1079-7114 \urlprefix\url{http://dx.doi.org/10.1103/PhysRevLett.130.236401}

\bibitem{vieijra2020}
Vieijra T, Casert C, Nys J, De~Neve W, Haegeman J, Ryckebusch J and Verstraete F 2020 {\em Phys. Rev. Lett.\/} {\bf 124}(9) 097201 \urlprefix\url{https://link.aps.org/doi/10.1103/PhysRevLett.124.097201}

\bibitem{vieijra2021}
Vieijra T and Nys J 2021 {\em Phys. Rev. B\/} {\bf 104}(4) 045123 \urlprefix\url{https://link.aps.org/doi/10.1103/PhysRevB.104.045123}

\bibitem{singh2012}
Singh S and Vidal G 2012 {\em Phys. Rev. B\/} {\bf 86}(19) 195114 \urlprefix\url{https://link.aps.org/doi/10.1103/PhysRevB.86.195114}

\bibitem{schmoll2020}
Schmoll P, Singh S, Rizzi M and Orús R 2020 {\em Annals of Physics\/} {\bf 419} 168232 ISSN 0003-4916 \urlprefix\url{http://dx.doi.org/10.1016/j.aop.2020.168232}

\bibitem{yang2022}
Yang J, Sandvik A~W and Wang L 2022 {\em Phys. Rev. B\/} {\bf 105}(6) L060409 \urlprefix\url{https://link.aps.org/doi/10.1103/PhysRevB.105.L060409}

\bibitem{gong2014}
Gong S~S, Zhu W, Sheng D~N, Motrunich O~I and Fisher M~P~A 2014 {\em Phys. Rev. Lett.\/} {\bf 113}(2) 027201 \urlprefix\url{https://link.aps.org/doi/10.1103/PhysRevLett.113.027201}

\bibitem{carleo2017}
Carleo G and Troyer M 2017 {\em Science\/} {\bf 355} 602–606 ISSN 1095-9203 \urlprefix\url{http://dx.doi.org/10.1126/science.aag2302}

\bibitem{nomura2021b}
Nomura Y 2021 {\em Journal of Physics: Condensed Matter\/} {\bf 33} 174003 ISSN 1361-648X \urlprefix\url{http://dx.doi.org/10.1088/1361-648X/abe268}

\bibitem{roth2021}
Roth C and MacDonald A~H 2021 Group convolutional neural networks improve quantum state accuracy (\textit{Preprint} \eprint{2104.05085})

\bibitem{roth2023}
Roth C, Szab\'o A and MacDonald A~H 2023 {\em Phys. Rev. B\/} {\bf 108}(5) 054410 \urlprefix\url{https://link.aps.org/doi/10.1103/PhysRevB.108.054410}

\bibitem{viteritti2024}
Viteritti L~L, Rende R, Parola A, Goldt S and Becca F 2024 Transformer wave function for the shastry-sutherland model: emergence of a spin-liquid phase (\textit{Preprint} \eprint{2311.16889})

\bibitem{rende2023}
Rende R, Viteritti L~L, Bardone L, Becca F and Goldt S 2024 {\em Communications Physics\/} {\bf 7} ISSN 2399-3650 \urlprefix\url{http://dx.doi.org/10.1038/s42005-024-01732-4}

\bibitem{hibat2020}
Hibat-Allah M, Ganahl M, Hayward L~E, Melko R~G and Carrasquilla J 2020 {\em Phys. Rev. Res.\/} {\bf 2}(2) 023358 \urlprefix\url{https://link.aps.org/doi/10.1103/PhysRevResearch.2.023358}

\bibitem{Lyu_2023}
Lyu C, Xu X, Yung M~H and Bayat A 2023 {\em Quantum\/} {\bf 7} 899 ISSN 2521-327X \urlprefix\url{http://dx.doi.org/10.22331/q-2023-01-19-899}

\bibitem{gard2020efficient}
Gard B~T, Zhu L, Barron G~S, Mayhall N~J, Economou S~E and Barnes E 2020 {\em npj Quantum Information\/} {\bf 6} 10

\bibitem{Seki2020}
Seki K, Shirakawa T and Yunoki S 2020 {\em Phys. Rev. A\/} {\bf 101}(5) 052340 \urlprefix\url{https://link.aps.org/doi/10.1103/PhysRevA.101.052340}

\bibitem{Meyer2023}
Meyer J~J, Mularski M, Gil-Fuster E, Mele A~A, Arzani F, Wilms A and Eisert J 2023 {\em PRX Quantum\/} {\bf 4} ISSN 2691-3399 \urlprefix\url{http://dx.doi.org/10.1103/PRXQuantum.4.010328}

\bibitem{Chang2023}
Chang S~Y, Grossi M, Saux B~L and Vallecorsa S 2023 Approximately equivariant quantum neural network for $p4m$ group symmetries in images (\textit{Preprint} \eprint{2310.02323})

\bibitem{Le2023}
Le I~N~M, Kiss O, Schuhmacher J, Tavernelli I and Tacchino F 2023 Symmetry-invariant quantum machine learning force fields (\textit{Preprint} \eprint{2311.11362})

\bibitem{cerezo2021variational}
Cerezo M, Arrasmith A, Babbush R, Benjamin S~C, Endo S, Fujii K, McClean J~R {\em et~al.\/} 2021 {\em Nature Reviews Physics\/} {\bf 3} 625--644

\bibitem{Kattemolle_2022}
Kattem\"olle J and van Wezel J 2022 {\em Phys. Rev. B\/} {\bf 106}(21) 214429 \urlprefix\url{https://link.aps.org/doi/10.1103/PhysRevB.106.214429}

\bibitem{mizusaki2004}
Mizusaki T and Imada M 2004 {\em Phys. Rev. B\/} {\bf 69}(12) 125110 \urlprefix\url{https://link.aps.org/doi/10.1103/PhysRevB.69.125110}

\bibitem{preskill2018quantum}
Preskill J 2018 {\em Quantum\/} {\bf 2} 79 ISSN 2521-327X \urlprefix\url{http://dx.doi.org/10.22331/q-2018-08-06-79}

\bibitem{Tüysüz2024}
Tüysüz C, Chang S~Y, Demidik M, Jansen K, Vallecorsa S and Grossi M 2024 Symmetry breaking in geometric quantum machine learning in the presence of noise (\textit{Preprint} \eprint{2401.10293})

\bibitem{PhysRevLett.119.180509}
Temme K, Bravyi S and Gambetta J~M 2017 {\em Phys. Rev. Lett.\/} {\bf 119}(18) 180509 \urlprefix\url{https://link.aps.org/doi/10.1103/PhysRevLett.119.180509}

\bibitem{wecker2015progress}
Wecker D, Hastings M~B and Troyer M 2015 {\em Physical Review A\/} {\bf 92} 042303

\bibitem{wiersema2020}
Wiersema R, Zhou C, de~Sereville Y, Carrasquilla J~F, Kim Y~B and Yuen H 2020 {\em PRX Quantum\/} {\bf 1} ISSN 2691-3399 \urlprefix\url{http://dx.doi.org/10.1103/PRXQuantum.1.020319}

\bibitem{mele2022}
Mele A~A, Mbeng G~B, Santoro G~E, Collura M and Torta P 2022 {\em Physical Review A\/} {\bf 106} ISSN 2469-9934 \urlprefix\url{http://dx.doi.org/10.1103/PhysRevA.106.L060401}

\bibitem{anselme2022}
Anselme~Martin B, Simon P and Rančić M~J 2022 {\em Physical Review Research\/} {\bf 4} ISSN 2643-1564 \urlprefix\url{http://dx.doi.org/10.1103/PhysRevResearch.4.023190}

\bibitem{wierichs2020}
Wierichs D, Gogolin C and Kastoryano M 2020 {\em Physical Review Research\/} {\bf 2} ISSN 2643-1564 \urlprefix\url{http://dx.doi.org/10.1103/PhysRevResearch.2.043246}

\bibitem{ho2019}
Ho W~W and Hsieh T~H 2019 {\em SciPost Physics\/} {\bf 6} ISSN 2542-4653 \urlprefix\url{http://dx.doi.org/10.21468/SciPostPhys.6.3.029}

\bibitem{feulner2022}
Feulner V and Hartmann M~J 2022 {\em Phys. Rev. B\/} {\bf 106}(14) 144426 \urlprefix\url{https://link.aps.org/doi/10.1103/PhysRevB.106.144426}

\bibitem{Vatan2004}
Vatan F and Williams C 2004 {\em Phys. Rev. A\/} {\bf 69}(3) 032315 \urlprefix\url{https://link.aps.org/doi/10.1103/PhysRevA.69.032315}

\bibitem{Childs2012}
Childs A~M and Wiebe N 2012 {\em Quantum Information and Computation\/} {\bf 12} ISSN 1533-7146 \urlprefix\url{http://dx.doi.org/10.26421/QIC12.11-12}

\bibitem{Berry2014}
Berry D~W, Childs A~M, Cleve R, Kothari R and Somma R~D 2014 Exponential improvement in precision for simulating sparse hamiltonians {\em Proceedings of the Forty-Sixth Annual ACM Symposium on Theory of Computing\/} STOC '14 (New York, NY, USA: Association for Computing Machinery) p 283–292 ISBN 9781450327107 \urlprefix\url{https://doi.org/10.1145/2591796.2591854}

\bibitem{Carrera_Vazquez_2023}
Carrera~Vazquez A, Egger D~J, Ochsner D and Woerner S 2023 {\em Quantum\/} {\bf 7} 1067 ISSN 2521-327X \urlprefix\url{http://dx.doi.org/10.22331/q-2023-07-25-1067}

\bibitem{Chakraborty2023}
Chakraborty S 2023 Implementing any linear combination of unitaries on intermediate-term quantum computers (\textit{Preprint} \eprint{2302.13555})

\bibitem{krantz2019quantum}
Krantz P, Kjaergaard M, Yan F, Orlando T~P, Gustavsson S and Oliver W~D 2019 {\em Applied Physics Reviews\/} {\bf 6} 021318 \urlprefix\url{https://pubs.aip.org/aip/apr/article/6/2/021318/570326/A-quantum-engineer-s-guide-to-superconducting}

\bibitem{breuer2002theory}
Breuer H~P, Petruccione F {\em et~al.\/} 2002 {\em The theory of open quantum systems\/} (Oxford University Press on Demand)

\bibitem{PhysRevA.104.062432}
Georgopoulos K, Emary C and Zuliani P 2021 {\em Phys. Rev. A\/} {\bf 104}(6) 062432 \urlprefix\url{https://link.aps.org/doi/10.1103/PhysRevA.104.062432}

\bibitem{dibartolomeo2023novel}
Di~Bartolomeo G, Vischi M, Cesa F, Wixinger R, Grossi M, Donadi S and Bassi A 2023 {\em Phys. Rev. Res.\/} {\bf 5}(4) 043210 \urlprefix\url{https://link.aps.org/doi/10.1103/PhysRevResearch.5.043210}

\bibitem{vischi2023simulating}
Vischi M, Di~Bartolomeo G, Proietti M, Koudia S, Cerocchi F, Dispenza M and Bassi A 2024 {\em Phys. Rev. Res.\/} {\bf 6}(3) 033337 \urlprefix\url{https://link.aps.org/doi/10.1103/PhysRevResearch.6.033337}

\bibitem{nielsen2000quantum}
Nielsen M~A and Chuang I~L 2000 {\em Quantum computing and quantum information\/} (Cambridge University Press, Cambridge)

\bibitem{benenti2019principles}
Benenti G, Casati G, Rossini D and Strini G 2019 {\em Principles of Quantum Computation and Information: A Comprehensive Textbook\/} (World Scientific)

\bibitem{Sarovar2020detectingcrosstalk}
Sarovar M, Proctor T, Rudinger K, Young K, Nielsen E and Blume-Kohout R 2020 {\em {Quantum}\/} {\bf 4} 321 ISSN 2521-327X \urlprefix\url{https://doi.org/10.22331/q-2020-09-11-321}

\bibitem{naghiloo2019introduction}
Naghiloo M 2019 Introduction to experimental quantum measurement with superconducting qubits (\textit{Preprint} \eprint{1904.09291})

\bibitem{van2022model}
Van Den~Berg E, Minev Z~K and Temme K 2022 {\em Physical Review A\/} {\bf 105} 032620

\bibitem{ibm_quantum_exp}
 2024 {Ibm quantum compute resources} \urlprefix\url{https://quantum-computing.ibm.com/services/resources}

\bibitem{unitaryfolding}
Li Y and Benjamin S~C 2017 {\em Phys. Rev. X\/} {\bf 7}(2) 021050 \urlprefix\url{https://link.aps.org/doi/10.1103/PhysRevX.7.021050}

\bibitem{Giurgica_Tiron_2020}
Giurgica-Tiron T, Hindy Y, LaRose R, Mari A and Zeng W~J 2020 Digital zero noise extrapolation for quantum error mitigation {\em 2020 IEEE International Conference on Quantum Computing and Engineering (QCE)\/} (IEEE) \urlprefix\url{http://dx.doi.org/10.1109/QCE49297.2020.00045}

\bibitem{pennylane2022}
Bergholm V, Izaac J, Schuld M, Gogolin C, Ahmed S and et~al V~A 2022 Pennylane: Automatic differentiation of hybrid quantum-classical computations (\textit{Preprint} \eprint{1811.04968})

\bibitem{lanczos1950iterative}
Lanczos C 1950 {\em J. Res. Nat. Bur. Standards\/} {\bf 45} 225--280

\bibitem{Ragone24}
Ragone M, Bakalov B~N, Sauvage F, Kemper A~F, Ortiz~Marrero C, Larocca M and Cerezo M 2024 {\em Nature Communications\/} {\bf 15} ISSN 2041-1723 \urlprefix\url{http://dx.doi.org/10.1038/s41467-024-49909-3}

\bibitem{Fontana24}
Fontana E, Herman D, Chakrabarti S, Kumar N, Yalovetzky R, Heredge J, Sureshbabu S~H and Pistoia M 2024 {\em Nature Communications\/} {\bf 15} ISSN 2041-1723 \urlprefix\url{http://dx.doi.org/10.1038/s41467-024-49910-w}

\bibitem{Zhang2022}
Zhang K, Liu L, Hsieh M~H and Tao D 2022 Escaping from the barren plateau via gaussian initializations in deep variational quantum circuits {\em Advances in Neural Information Processing Systems\/} ed Oh A~H, Agarwal A, Belgrave D and Cho K \urlprefix\url{https://openreview.net/forum?id=jXgbJdQ2YIy}

\bibitem{astrakhantsev2023}
Astrakhantsev N, Mazzola G, Tavernelli I and Carleo G 2023 {\em Phys. Rev. Res.\/} {\bf 5}(3) 033225 \urlprefix\url{https://link.aps.org/doi/10.1103/PhysRevResearch.5.033225}

\bibitem{dur2000}
D\"ur W, Vidal G and Cirac J~I 2000 {\em Phys. Rev. A\/} {\bf 62}(6) 062314 \urlprefix\url{https://link.aps.org/doi/10.1103/PhysRevA.62.062314}

\bibitem{mcclung2020constructions}
McClung J 2020 {\em Worcester Polytechnic Institute\/}

\bibitem{Majumdar1969}
{Majumdar} C~K and {Ghosh} D~K 1969 {\em Journal of Mathematical Physics\/} {\bf 10} 1388--1398

\bibitem{manzano2020short}
Manzano D 2020 {\em Aip Advances\/} {\bf 10}

\end{thebibliography}

\end{document}